\documentstyle[aps,prb,tighten,eqsecnum]{revtex}

\begin{document}

\title{The sn-pole approximation in the Composite Operator Method}
\author{F. Mancini\cite{Emancini}}
\address{Dipartimento di Scienze Fisiche ``E. R. Caianiello`` - Unit\`a INFM di Salerno\\
Universit\`a degli Studi di Salerno, 84081 Baronissi (SA), Italy}
\date{July 21, 2000}
\maketitle
\begin{abstract}
A well-established method to deal with highly correlated systems is based on
the expansion of the Green's function in terms of spectral moments. In the
context of the Composite Operator Method one approximation is proposed: a set
of $n$ composite fields is assumed as fundamental basis and the dynamics is
considered up to the order $s$. The resulting Green's function has a $sn$-pole
structure. The truncation of the hierarchy of the equations of motion is made
at the $s$-th order and the first $s-1$ equations are treated exactly. A
theorem, which rules the conservation of the spectral moments, is presented.
The procedure is applied to the Hubbard model and a recurrence relation for the
calculation of its electronic spectral moments is derived.
\end{abstract}

\section{Introduction}
The discovery of new materials with unusual properties \cite{one} requires the
formulation of new schemes of calculation capable to catch the essential
physics of correlations among particles. One powerful tool is the Green's
function formalism. Most of the properties of a many-body system can be fully
described in terms of the thermal single-particle Green's function. For finite
systems the spectral function $\rho ({\bf k},\omega )$ can always be written
as an infinite weighted sum of delta functions and the retarded Green's
function \cite{two} can be expressed as
\begin{equation}
g({\bf k},\omega )=\int\limits_{-\infty }^{+\infty } d\omega' \frac{\rho ({\bf
k},\omega' )}{\omega -\omega' } =\sum\limits_{i=1}^{\infty } \frac{\sigma
^{(i)}
({\bf k})}{%
\omega -E_{i} ({\bf k})+i\eta }
\end{equation}   
where $E_{i} ({\bf k})$ and $\sigma ^{(i)} ({\bf k})$ are the energy spectrum
and the spectral density function, respectively, of the $i$-th elementary
excitation. In the way of developing approximate methods, one line of thinking
is to approximate the Green's function by a sum on a finite number of
excitations (pole approximation). The main question that arises is how the
poles must be chosen (i.e., which excitations do describe the physics we are
interested to analyze?). Let us consider a Hamiltonian $H=H[\varphi]$, where
$\varphi $ denotes a field operator. In the spectral density approach ({\em
SDA}) \cite{three} one truncates the sum in (1.1) by picking up the first $s$
terms ($s$-pole approximation). The unknown functions $E_{i} ({\bf k})$ and
$\sigma ^{(i)} ({\bf k})$ are fixed by the set of $2s$ nonlinear equations
\begin{equation}
 m^{(k)} ({\bf k})=\sum\limits_{i=1}^{s} \/E_{i}^{k} ({\bf k})\sigma
^{(i)} ({\bf k})
\end{equation}   
where $m^{(k)} ({\bf k})$, the moments of the spectral density, are calculated
by means of the equations of motion:
\begin{equation}
m^{(k)} ({\bf k})=F.T.\left[ \left( i\partial /\partial t_{x} \right) ^{k}
<\lbrack \varphi (x),\varphi ^{\dagger} (y)]_{\pm } >\right] _{t_{x} =t_{y} }
\end{equation}   
Eq.~(1.2) is derived by making use of the exact relation (conservation of
spectral moments)
\begin{equation}
m^{(k)} ({\bf k})=\int\limits_{-\infty }^{+\infty } d\omega \,\rho ({\bf
k},\omega )\omega ^{k}
\end{equation}   
The symbol $\lbrack \cdots ]_{\pm } $ denotes the anticommutator ($+$) or the
commutator ($-$) according to the statistics the field $\varphi$ obeys; the
symbol $<\cdots>$ denotes the thermal average on the grand canonical ensemble;
$F.T.$ indicates the Fourier transform. This procedure, which is an expansion
of the Green's function in terms of the first $2s$ spectral moments, gives a
good description in the high-energy region but certainly fails in the
low-energy region, where one needs $s$ to be very large.

When we consider systems where the correlation is very high the procedure
above illustrated may not be adequate. Due to the strong interactions the
original particles are not observable; new excitations appear and it is more
convenient to describe the system in terms of new field operators. Collective
behaviors in forms of bound states, resonances and diffused modes emerge as
the physical fields. Although some of them are not stable excitations, they
give considerable contributions to the physical processes and therefore it is
sometimes convenient to promote them to the role of well-defined quasi-particle
excitations. The choice of new fundamental particles, whose properties have to
be self-consistently determined by dynamics, symmetries and boundary
conditions, becomes relevant. The necessity of developing a formulation to
treat composite excitations has seen remarkable developments in the many-body
theory in the form of an assortment of techniques that may be termed composite
particle methods. The projection method \cite{four,one}, the method of equation
of motion \cite{five}, the slave boson method \cite{six}, the spectral density
approach \cite{three}, the composite operator method ({\em COM})
\cite{seven,eight,nine} are all along this line. In the {\em COM} language let
us write $H=H_{I}+H_{II}$, where $H_{I} $ is chosen according to the physics of
the problem under investigation,
and let $%
\psi =\{\psi _{i} \}\;\lbrack i=1,\cdots ,n]$ be a complete set of
eigenoperators of $H_{I}$
\begin{equation}
i\frac{\partial }{\partial t} \psi (x)=\lbrack \psi (x),H_{I} ]=\epsilon _{I}
(-i\vec{\nabla} )\psi (x)
\end{equation}   
The fields $\psi _{n} $ are generally constructed as functionals of the
original field $\varphi $ and are called composite fields. Then, $H_{II} $
will be taken into account by some approximate methods that will use $\psi $
as operatorial basis. Various approximations have been used to estimate the
contribution coming from $H_{II} $. One method, the $n-$ pole approximation,
linearizes the full equation of motion
\begin{equation}
i\frac{\partial }{\partial t} \psi (x)=\lbrack \psi (x),H]=\epsilon (-i
\vec{\nabla})\psi (x)+\delta j(x)\approx \epsilon (-i \vec{\nabla})\psi (x)
\end{equation}   
where the eigenvalue or energy matrix $\epsilon $ is self-consistently
calculated by means of the equation
\begin{equation}
\epsilon ({\bf k})M^{(0)} ({\bf k})=M^{(1)} ({\bf k})
\end{equation}   
$M^{(k)} ({\bf k})$ is the generalized spectral moment, given by the
$n\/\times \/n $ matrix
\begin{equation}
M^{(k)} ({\bf k})=F.T.\left[ \left( i\partial /\partial t_{x} \right) ^{k}
<\lbrack \psi (x),\psi ^{\dagger} (y)]_{\pm } >\right] _{t_{x} =t_{y} }
\end{equation}   
The decomposition in (1.6) is chosen in such a way that the higher-order
source $\delta j(x)$ does not have a projection on the fundamental basis $\{
\psi _{i}\}$ . In this approximation the retarded ``single-particle`` Green's
function $G({\bf k},\omega )=F.T.<R\lbrack \psi (x)\psi ^{\dagger} (y)]>$ has
a $n$-pole structure, where the number of poles is equal to the number of
composite fields. It is clear that the properties of the Green's function are
controlled by the dynamics and by the choice of the new basis. When this
approximation is used the {\em COM} has some similarity with the {\em SDA}:
both approaches correspond to expanding the Green's function in terms of
spectral moments. However, there is a fundamental difference. In the {\em SDA}
there is no freedom in choosing the spectral moments, which are completely
fixed by the equations of motion. At the contrary, in the {\em COM} the
spectral moments are mainly determined by the choice of the operatorial basis.
The idea is that in the framework of approximating the Green's function by a
finite number of excitations, a proper choice of the basis might give a better
convergence and a more accurate description of the low-energy region. Let us
write
\begin{equation}
M^{(k)} ({\bf k})=\tilde{M} ^{(k)} ({\bf k})+\delta M^{(k)} ({\bf k})
\end{equation}   
where
\begin{equation}
\tilde{M} ^{(k)} ({\bf k})=\epsilon ^{k} ({\bf k})M^{(0)} ({\bf k})
\end{equation}   
\begin{equation}
\delta M^{(k)} ({\bf k})=\sum\limits_{m=1}^{k-1} F.T.\epsilon ^{k-m-1} \left(
i\partial /\partial t_{x} \right) ^{m} <\left [ \delta j(x),\psi ^{\dagger}
(y)\right ]_{\pm } >_{t_{x} =t_{y} }
\end{equation}   
$\tilde{M} ^{(k)} ({\bf k})$ corresponds to the spectral moment calculated in
the n-pole approximation, where the higher-order field $\delta j(x)$ in the
Heisenberg equation (1.6) is neglected. In Ref.~\onlinecite{ten} we proved the
following theorems. \\

\noindent {\bf Theorem I} Given a set of composite fields $\{\psi _{i}
,\;i=1,....,n\}$, satisfying linearized Heisenberg equations $i(\partial
/\partial t)\psi =\epsilon \psi $, the approximate spectral moments $\tilde{M}
^{(k)}
({\bf k})$, are conserved at any order. \\

\noindent {\bf Theorem II} Given a set of composite fields $\{\psi _{l}
,\;l=1,....,n\}$ , if the subset $\{\psi _{l} ,\;l=1,....,n-1\}$ satisfies
linear Heisenberg equations $i\frac{\partial }{\partial t} \psi _{l}
(x)=\sum\limits_{p=1}^{l+1} \gamma _{lp} (-i\vec{\nabla} )\psi _{p} (x)$, then
the first $2(n-l+1)$ spectral moments for the field $\psi _{l} $ [that is
$M_{ll}^{(k)} ({\bf k})$ ] are conserved. \\

The Theorem II shows that the {\em SDA} corresponds to {\em COM} when a
particular choice of the basic fields is considered and a pole approximation
is used. In {\em COM} we have more flexibility, in the sense that there is no
restriction on the choice of the basic fields.

A different approximation has been proposed in Refs.~\onlinecite{eleven} and
\onlinecite{twelve}, where in the context of the $t$-$J$ and Hubbard models the
incoherent part of the Green's function has been taken into account by a
two-site approximation in combined use of the resolvent method. Although this
approximation gave excellent results, the numerical calculations are very heavy
in the small temperature regime. A pole approximation is much more convenient
for practical calculations and there are no restrictions on the range of
physical parameters. As shown in a variety of works, the pole approximation
gives a satisfactory account of the local and spectral features. Therefore, it
is interesting to explore the possibility to improve the approximation of the
{\em COM} by improving the $n$-pole approximation. In this article we propose
a new approximation, the $sn$-pole approximation, where we combine two
ingredients: the choice of the fields and the dynamics. The procedure is the
following: (i) we fix $n$ basic fields by some reasonable criterion based on
the physical properties of the model; (ii) we take into account the dynamics
up to the $s$-th order. In this scheme the Green's function has a $sn$-pole
structure
\begin{equation}
G({\bf k},\omega )=\sum\limits_{i=1}^{sn} \frac{\sigma ^{(i)} ({\bf
k})}{\omega -E_{i} ({\bf k})+i\eta }
\end{equation}   
where $E_{i} ({\bf k})$ and $\sigma ^{(i)} ({\bf k})$ are expressed in terms
of the $2s$ spectral moment matrices $M^{({\bf k})} (k)\quad [ 0\leq k\leq 2s-1]$. Let $%
A^{(m)} ({\bf k})\;\lbrack m=1,\ldots ,s-1]$ be matrices of rank $n\times n$
determined by the equations
\begin{equation}
M^{(s+k)} ({\bf k})-\sum\limits_{m=0}^{s-1} A^{(m)} ({\bf k})M^{(m+k)} ({\bf
k})=0\quad \quad \quad \left[ {\rm for}\quad 0\leq k\leq s-1\right ]
\end{equation}   
Then, the main results of this work can be summarized in the following
theorems. \\

\noindent {\bf Theorem III} Given a set of composite fields
$\{\psi_{i},\;i=1,\ldots,n\}$, satisfying linearized Heisenberg equations
\begin{equation}
\left( i\frac{\partial }{\partial t} \right) ^{s} \psi
(x)-\sum\limits_{m=0}^{s-1} A^{(m)} (-i\vec{\nabla} )\left( i\frac{\partial }{
\partial t} \right) ^{m} \psi (x)=0
\end{equation}   
the first $2s$ generalized spectral moments are conserved and the approximate
spectral moments $\tilde{M}^{(k)}({\bf k})$ are conserved at any order. \\

\noindent {\bf Theorem IV} Given a set of composite fields $\{\psi _{l}
,\;l=1,\ldots,n\}$, if the subset $\{\psi _{l} ,\;l=1,....,n-1\}$ satisfies
the Heisenberg equations (1.14), the first $2s(n-l+1)$ spectral moments for the
field $\psi _{l} (x)$ [that is
$M_{ll}^{(k)}({\bf k})$] are conserved. \\

When a particular choice of the basic fields is considered, we will show that
the $sn$-pole approximation is equivalent to {\em SDA} with $s\rightarrow sn$.

\section{Equations of motion and Green's function}

The composite field $\psi (x)$ satisfies the equation of motion
\begin{equation}
i\frac{\partial }{\partial t} \psi (x)=\lbrack \psi (x),H]=j^{(1)} (x)
\end{equation}   
The source $j^{(1)} (x)$ is expressed, in general, in terms of higher-order
composite fields with a dynamics determined by the Heisenberg equation
\begin{equation}
i\frac{\partial }{\partial t} j^{(1)} (x)=\lbrack j^{(1)} (x),H]=j^{(2)} (x)
\end{equation}   
The process can be continued and, in general, we have for the $s$-th derivative
the equation
\begin{equation}
\left( i\frac{\partial }{\partial t} \right) ^{s} \psi (x)=j^{(s)} (x)\quad
\quad \quad \lbrack for\quad s\geq 1]
\end{equation}   
Let us rewrite Eq.~(2.3) under the form
\begin{equation}
\left( i\frac{\partial }{\partial t} \right) ^{s} \psi
(x)-\sum\limits_{m=0}^{s-1} A^{(m)} (-i\vec{\nabla} )\left( i\frac{\partial
}{%
\partial t} \right) ^{m} \psi (x)=\delta j^{(s)} (x)
\end{equation}   
where $A^{(m)} (-i\vec{\nabla} )$ are matrices of rank $n$ determined by the
equations
\begin{equation}
<\lbrack \delta j^{(s)} ({\bf x},t),\left( -i\frac{\partial }{\partial t}
\right) ^{k} \psi ^{\dagger} ({\bf y},t)]_{\pm } >=0\quad \quad \quad \lbrack
for\quad 0\leq k\leq s-1]
\end{equation}   
By substituting (2.4) into (2.5) we find that the matrices $A^{(m)} ({\bf
k})\quad \lbrack 0\leq m\leq s-1]$ are determined by the following system of
equations
\begin{equation}
M^{(s+k)} ({\bf k})-\sum\limits_{m=0}^{s-1} A^{(m)} ({\bf k})M^{(m+k)} ({\bf
k})=0\quad \quad \quad \lbrack for\quad 0\leq k\leq s-1]
\end{equation}   
The evaluation of the $s$ matrices $A^{(m)} ({\bf k})$ requires the knowledge
of the first $2s$ generalized spectral moments $M^{(k)} ({\bf k})$, that can
be calculated by means of (1.8). By using the equation of motion (2.4) the
retarded Green's function $G(x,y)=<R\lbrack \psi (x)\psi ^{\dagger} (y)]>$
satisfies the equation
\begin{eqnarray}
&&\sum\limits_{k=0}^{s} A^{(k)}(-i\vec{\nabla} _{x} )\left( i\frac{
\partial}{\partial t_{x} }
\right) ^{k} G(x,y) \nonumber\\
&& \quad \quad =\sum\limits_{k=1}^{s} (i)^{k} \left( \frac{ \partial}{\partial
t_{x} } \right) ^{k-1} \delta (t_{x} -t_{y} )\sum\limits_{m=k}^{s} A^{(m)}
(-i\vec{\nabla} _{x}
)I^{(m-k)} ({\bf x},{\bf y})  \\
&& \quad \quad -<R\lbrack \delta j^{(s)} (x)\psi ^{\dagger} (y)]> \nonumber
\end{eqnarray}          
where
\begin{equation}
I^{(k)} ({\bf x},{\bf y})=<[j^{(k)} (x),\psi ^{\dagger} (y)]_{\pm}>_{t_{x}
=t_{y} }
\end{equation}      
with the convention that
\begin{equation}
A^{(s)} (-i\vec{\nabla} )=-1\end{equation}      
Eq.~(2.7) is an exact equation. We shall define our approximation, $sn$-pole
approximation, by neglecting the term $<R\lbrack \delta
j^{(s)}(x)\psi^{\dagger} (y)]>$ , with $\delta j^{(s)} (x)$ fixed by the
requirement of vanishing projections on the basis $[j^{(k)}(x);\quad 0\leq
k\leq s-1]$ with the convection $j^{(0)}(x)=\psi(x)$. Then, the solution of
(2.7) in momentum space is
\begin{equation}
G({\bf k},\omega )=\frac{1}{\omega ^{s} -\sum\limits_{k=0}^{s-1} A^{(k)} ({\bf
k})\omega ^{k} } \sum\limits_{k=1}^{s} C^{(s,k)} ({\bf k})\omega ^{k-1}
\end{equation}      
where we defined
\begin{equation}
C^{(s,k)} ({\bf k})=-\sum\limits_{m=k}^{s} A^{(m)} ({\bf k})M^{(m-k)} ({\bf
k})\quad \quad
\quad \lbrack for\quad 1\leq k\leq s]\end{equation}      
Eq.~(2.10) shows that $G({\bf k},\omega )$ is expressed in terms of the first
$2s$ generalized spectral moments. This expansion is dictated by the choice of
the composite field and by the dynamics up to the $s$-th order. Expression
(2.10) can be rewritten in the spectral form
\begin{equation}
G({\bf k},\omega )=\sum\limits_{i=1}^{sn} \frac{\sigma ^{(i)} ({\bf k})}{\omega
-E_{i}
({\bf k})+i\eta } \end{equation}      
where the energy spectra $E_{i} ({\bf k})$ are the roots of the equation
\begin{equation}
\sum\limits_{m=0}^{sn} \omega ^{m} a_{m} ({\bf k})=0\end{equation}
with coefficients $a_{m} ({\bf k})$ defined by
\begin{equation}
\begin{tabular}{l}
$a_{m} ({\bf k})=\frac{1}{m!} \left[ \frac{\partial }{\partial \omega ^{m} }
Det \{\omega ^{s} -\sum\limits_{k=0}^{s-1} A^{(k)} ({\bf k})\omega ^{k} \}
\right]
_{\omega =0} \quad \quad \lbrack for\quad 0\leq m\leq sn-1] $ \\
$a_{sn} ({\bf k})=1 $%
\end{tabular}
\end{equation}     
The spectral density functions $\sigma ^{(i)} ({\bf k})$ are expressed as
\begin{equation}
\sigma ^{(i)} ({\bf k})=\frac{1}{b^{(i)} ({\bf k})} \sum\limits_{m=0}^{sn-1}
E_{i}^{m}
({\bf k})\lambda ^{(m)} ({\bf k})\end{equation}      
where we put
\begin{equation}
b^{(i)} ({\bf k})= \mathop{\displaystyle \prod } \limits_{j=1,\,j\neq i}^{sn}
\lbrack E_{i} ({\bf k})-E_{j}
({\bf k})]\end{equation}      
and the $\lambda ^{(m)} ({\bf k})$ are $n \times n$ matrices, determined by the
recurrence relation
\begin{equation}
\begin{tabular}{l}
$\lambda ^{(k)} ({\bf k})=\sum\limits_{p=0}^{s-1} A^{(p)} ({\bf k})\lambda
^{(k+s-p)}({\bf k}) $ \\
\quad \quad \quad $-\sum\limits_{p=0}^{s-1} \sum\limits_{m=p+1}^{s}
a_{k+s-p} ({\bf k})A^{(m)} ({\bf k})M^{(m-p-1)} ({\bf k}) $%
\end{tabular}
\quad \quad \quad [ {\rm for}\quad 0\leq k\leq sn-1]
\end{equation}      
with the convention
\begin{equation}
\begin{tabular}{l}
$\lambda ^{(m)} ({\bf k})=0\quad \quad for\quad m\notin \lbrack 0,sn-1] $\\
$a_{m} ({\bf k})=0\quad \quad for\quad m\notin \lbrack 0,sn] $%
\end{tabular}
\end{equation}        
When $k\geq sn-2s$ the matrices $\lambda ^{(k)} ({\bf k})$ can be simply
expressed in terms of the spectral moments by means of the relation
\begin{equation}
\lambda ^{(k)} ({\bf k})=\sum\limits_{m=0}^{sn-k-1} a_{sn-m} ({\bf
k})M^{(sn-k-1-m)} ({\bf k})\quad \quad \quad \lbrack for\quad sn-2s\leq k\leq
sn-1]
\end{equation}      
In particular, we see that for $n\leq 2$ the relation (2.19) is valid for all
values of $s$ and for $k$ varying in the all interval $0\leq k\leq sn-1 $.

Summarizing, the calculation of the Green's function in the $sn$-approximation
requires the following steps:
\begin{itemize}
\item (i) given a Hamiltonian, we choose a set $\{\psi (x)\}$ of $n$
composite field;
\item (ii) we calculate the first $2s$ generalized spectral moments
$M^{(k)} ({\bf k})$ by means of the formula (1.8);
\item (iii) we calculate the
matrices $A^{(m)} ({\bf k})\quad \lbrack 0\leq m\leq s-1]$ by solving the
linear system (2.6);
\item (iv) we calculate the ``characteristic`` coefficients
$a_{m} ({\bf k})\quad \lbrack 0\leq m\leq sn]$ by means of (2.14);
\item (v) the
energy spectra $E_{i} ({\bf k})\quad \lbrack 1\leq i\leq sn]$ are calculated
as the roots of the equation (2.13);
\item (vi) the matrices $\lambda ^{(m)} ({\bf
k})\quad \lbrack 0\leq m\leq sn-1]$ are calculated by means of the recurrence
relation (2.17);
\item (vi) the spectral functions $\sigma ^{(k)} ({\bf k})\quad
\lbrack 1\leq k\leq sn]$ are calculated by means of the expression (2.15).
\end{itemize}

In closing this section, it is worth noting that the truncation the hierarchy
of the equations of motion at the order $s$ implies that the first $s-1$
equations are treated exactly.

\section{Conservation of the spectral moments}

The spectral moments play an essential role in the formulation developed in
the previous Section. It is important to verify to which extent the exact
relation (1.4) is conserved in our approximation. In the $sn$-pole
approximation the spectral function has the expression
\begin{equation}
\rho ({\bf k},\omega )=\sum\limits_{i=1}^{sn} \sigma ^{(i)} ({\bf k})\delta
\lbrack
\omega -E_{i} ({\bf k})]\end{equation}      
Let us denote by $B^{(k)} ({\bf k})$ the spectral moments calculated in the
present approximation
\begin{equation}
B^{(k)} ({\bf k})=\int\limits_{-\infty }^{+\infty } d\omega \,\omega ^{k} \rho
({\bf k},\omega )=\sum\limits_{i=1}^{sn} E_{i}^{k} ({\bf k})\sigma ^{(i)}
({\bf k})\end{equation}      
The point to discuss is the relation between $B^{(k)} ({\bf k})$ and the exact
spectral moments $M^{(k)} ({\bf k})$ , calculated by means of (1.8). We shall
study separately the two quantities. In Appendix A we prove that the matrices
$B^{(k)} ({\bf k})$ satisfy the following relations
\begin{itemize}
  \item (i)For n=1
\begin{equation}
\begin{tabular}{l}
$B^{(k)} ({\bf k})=M^{(k)} ({\bf k})\quad \quad \quad \quad \quad \quad \quad
\quad
\quad for\quad 0\leq k\leq 2s-1 $ \\
$B^{(k)} ({\bf k})=-\sum\limits_{m=0}^{s-1} a_{m} ({\bf k})B^{(k-s+m)} ({\bf
k})\quad \quad
\quad \; for \quad k\geq s $%
\end{tabular}
\end{equation}         
  \item (ii) For $n\geq 2$
\begin{equation}
\begin{array}{ll}
B^{(k)} ({\bf k})= M^{(k)} ({\bf k})\qquad &{\rm for}\quad 0\leq k\leq 2s-1 \\
B^{(k)} ({\bf k})=\sum\limits_{p=0}^{s-1} A^{(p)} ({\bf k})B^{(k-s+p)} ({\bf
k})\qquad
&{\rm for}\ s\leq k\leq sn-1 \\
B^{(k)} ({\bf k})=-\sum\limits_{m=0}^{sn-1} a_{m}({\bf k})B^{(k-sn+m)} ({\bf
k})\qquad
&{\rm for}\ k\geq sn  \\
\end{array}
\end{equation}       
\end{itemize}

We see that the first $2s$ spectral moments are conserved. This is true for
any element of the matrix $M^{(k)}({\bf k})$, and therefore for all the fields
$\psi _{l} \;\lbrack 1\leq l\leq n]$. In Appendix B we show that the spectral
moments for $k\geq s$ can be written as
\begin{equation}
M^{(k)} ({\bf k})=\tilde{M} ^{(k)} ({\bf k})+\delta M^{(k)}
({\bf k})\end{equation}      
where $\tilde{M} ^{(k)} ({\bf k})$ and $\delta M^{(k)} ({\bf k})$ satisfy the
recurrence relations
\begin{equation}
\tilde{M} ^{(k)} ({\bf k})=\sum\limits_{m=0}^{s-1} A^{(m)} ({\bf k})\tilde{M}
^{(m+k-s)} ({\bf k})\quad \quad \quad \lbrack for\quad k\geq
s]\end{equation}      
\begin{equation}
\delta M^{(k)} ({\bf k})=F^{(k-s)} ({\bf k})+\sum\limits_{m=0}^{s-1} A^{(m)}
({\bf k})\delta M^{(m+k-s)} ({\bf k})\quad \quad \quad \lbrack for\quad k\geq
2s]\end{equation}      
with
\begin{equation}
F^{(k)} ({\bf k})=F.T.<[\left( i\frac{\partial }{\partial t} \right) ^{k}
\delta j^{(s)} (x),\psi ^{\dagger} (y)]_{\pm}>_{t_{x} =t_{y} }
\end{equation}     
Use of time translational invariance and of the constrains (2.5) leads to
\begin{equation}
\delta M^{(k)} ({\bf k})=0\quad \quad \quad \lbrack for\;0\leq k\leq
2s-1]\end{equation}      
We note that $\tilde{M} ^{(k)} (k)$ represents the spectral moment in the
approximation considered, when the higher-order field $\delta \/j^{(s)} $ is
neglected. Furthermore, in Appendix B we show that the matrices
$\tilde{M}^{(k)} ({\bf k})$ satisfy the recurrence relation
\begin{equation}
\tilde{M} ^{(k)} ({\bf k})=-\sum\limits_{m=0}^{sn-1} a_{m} ({\bf k})\tilde{M}
^{(k-sn+m)} ({\bf k})\quad \quad \quad \lbrack for\;k\geq
sn]\end{equation}      
By comparing (3.10) and (3.3)-(3.4) we see that for $k\geq sn$ $\tilde{M}
^{(k)} ({\bf k})$ and $B^{(k)} ({\bf k})$ satisfy the same recurrence
relation. But, we have shown that for $k\leq sn-1$
\begin{equation}
B^{(k)} ({\bf k})=\tilde{M} ^{(k)} ({\bf k})\end{equation}      
therefore the two matrices coincide at any order. This shows an internal
consistency of the approximation, in the sense that the exact relation (1.4)
is preserved at any order when calculated on the same ground. The above
discussion can be summarized in the Theorem III stated in Section I.

\section{Special choice of the basis}

Up to now the basic set ${\psi (x)}$ has not been specified. Let us consider
the case where the $n$ fields are chosen as
\begin{equation}
\left( i\frac{\partial }{\partial t} \right) ^{s} \psi _{l}
(x)=\sum\limits_{p=1}^{l+1} \tilde{A} _{lp}^{(0)} (-i\vec{\nabla} )\psi _{p}
(x)+\sum\limits_{m=1}^{s-1} \sum\limits_{p=1}^{l} A_{lp}^{(m)} (-i\vec{\nabla}
)\left( i\frac{\partial }{\partial t} \right) ^{m} \psi _{p} (x)
\end{equation}      
where $\tilde{A} ^{(0)} $ is a matrix of $n$ rows and $n+1$ columns and
$A^{(m)} \;\lbrack 1\leq m\leq s-1]$ are $n\times n$ matrices. That is, we
choose the first field $\psi _{1} (x)$ ; then, all the other fields $\psi _{l}
\quad \lbrack 2\leq l\leq n]$ are determined by means of the equation
\begin{eqnarray}
\tilde A_{l-1,l}^{(0)} (-i\vec{\nabla} )\psi _{l} (x)&=& \left(
i\frac{\partial}{\partial t} \right) ^{s} \psi _{l-1}
(x)-\sum\limits_{p=1}^{l-1} \tilde A_{l-1,p}^{(0)} (-i\vec{\nabla} )\left(
i\frac{\partial}{\partial t} \right)^{m}\psi_{p}(x)\nonumber\\
&& -\sum\limits_{m=1}^{s-1} \sum\limits_{p=1}^{l-1} A_{l-1,p}^{(m)}
(-i\vec{\nabla} )\left( i\frac{\partial }{\partial t} \right) ^{m} \psi _{p}
(x)
\end{eqnarray}          
It is clear that our basis satisfies the equation of motion
\begin{equation}
\left( i\frac{\partial }{\partial t} \right) ^{s} \psi
(x)=\sum\limits_{m=0}^{s-1} A^{(m)} (-i\vec{\nabla} )\left( i\frac{\partial
}{%
\partial t} \right) ^{m} \psi (x)+\delta j^{(s)}
(x)\end{equation}      
where $A^{(0)} $ is the square matrix obtained from $\tilde{A} ^{(0)} $ by
taking the first $n$ rows and $n$ columns. The matrices $\tilde{A} ^{(0)} $ ,
$A^{(m)} \;\lbrack 1\leq m\leq s-1]$ and the vector $\delta j_{l}^{(s)} $ have
the structure
\begin{equation}
\begin{tabular}{l}
$\tilde{A} \/_{l,p}^{(0)} =0\quad \quad for\quad
\begin{array}{l}
1\leq l\leq n-1 \\
l+2\leq p\leq n+1
\end{array}
$ \\
$A\/_{l,p}^{(m)} =0\quad \quad for\quad
\begin{array}{l}
1\leq l\leq n-1 \\
l+1\leq p\leq n
\end{array}
$ \\
$\delta j_{l}^{(s)} =0\quad \quad for\quad l\leq n-1 $%
\end{tabular}
\end{equation}            
We shall take the matrices $A^{(m)} \;\lbrack 1\leq m\leq s-1]$ in such a way
to satisfy the constrain (2.8). We have seen in Section 3 (cfr. Eq.~(3.5))
that the spectral moments can be written as $M^{(k)} ({\bf k})=\tilde{M} ^{(k)}
({\bf k})+\delta M^{(k)} ({\bf k})$ . For the special choice of the basis
considered in this Section, it is possible to show that the elements of the
matrices $F^{(k)} ({\bf k})$ , defined by Eq.~(3.8), satisfy the following
rules
\begin{equation}
\begin{tabular}{l}
$F_{l,p}^{(k)} ({\bf k})=0\quad \quad for\;l\leq n-1\quad \lbrack
for\;any\;value\;of\;p\;and\;k] $ \\
$F_{n,p}^{(k)} ({\bf k})=0\quad \quad for\;k\leq s(n-p+1)-1 $%
\end{tabular}
\end{equation}  
Then, the elements of the matrix $\delta M^{(k)} ({\bf k})$ satisfy the
following rule
\begin{equation}
\delta M_{ll}^{(k)} (k)=0\quad \quad for\quad k\leq
2s(n-l+1)-1\end{equation}      
By recalling the result $B^{(k)} ({\bf k})=\tilde{M} ^{(k)} ({\bf k})$ , we
have thus proved the Theorem IV, stated in Section I.

\section{Some examples}

In order to illustrate the formulation let us consider some specific examples.

\noindent {\it\bf Case $n=1$}

The solution of the system (1.17) is
\begin{equation}
A^{(k)} ({\bf k})=\sum\limits_{l=1}^{s} \left( \frac{1}{D({\bf k})} \right)
_{k+1,l}M^{(s+l-1)} ({\bf k})\quad \quad \quad \lbrack for\quad 0\leq k\leq
s-1]\end{equation}      
where $D({\bf k})$ is the $s\/\times \/s$ matrix
\begin{equation}
D({\bf k})=\left(
\begin{array}{ccc}
M^{(0)} & \cdots & M^{(s-1)} \\
\vdots & \cdots & \vdots \\
M^{(s-1)} & \cdots & M^{(2s-2)}
\end{array}
\right) \end{equation}      
Then, the retarded Green's function is given by
\begin{equation}
G({\bf k},\omega )=\sum\limits_{i=1}^{s} \frac{\sigma ^{(i)} ({\bf k})}{\omega
-E_{i}({\bf k})+i\eta } \end{equation}      
The energy spectra $E_{i} ({\bf k})$ are determined as the roots of the
equation (2.13), with characteristic coefficients $a_{m} ({\bf k})=-A^{(m)}
({\bf k})$ . The spectral functions are given by (2.15) with
\begin{equation}
\lambda ^{(k)} ({\bf k})=-\sum\limits_{m=0}^{s-k-1} A^{(s-m)} ({\bf
k})M^{(s-k-1-m)}({\bf k})\quad \quad \quad \lbrack for\quad 0\leq k\leq
s-1]\end{equation}      
This case is exactly equivalent to {\em SDA} and the previous formulas give a
complete solution to the problem of solving the system (1.2) for any value of
$s$.

\noindent {\it\bf Case $n=2$}

For $n=2$ we introduce the basic field
\begin{equation}
\psi (x)=\left(
\begin{array}{c}
\psi _{1} (x) \\
\psi _{2} (x)
\end{array}
\right) \end{equation}      
and consider the particular case where the second field is taken according to
the special choice (4.1). By noting that the equation of motion for the field
$\psi _{2} (x)$ is
\begin{equation}
\left( i\frac{\partial }{\partial t} \right) ^{k} \psi _{2} (x)=-\lbrack
A_{1,2}^{(0)} (-i\vec{\nabla} )]^{-1} \sum\limits_{m=0}^{s} A_{11}^{(m)}
(-i\vec{\nabla}
)j_{1}^{(k+m)} (x)\end{equation}      
the generalized spectral moments are given by
\begin{equation}
\begin{array}{l}
M_{11}^{(k)} ({\bf k})=m^{(k)} ({\bf k}) \\
M_{12}^{(k)} ({\bf k})=M_{21}^{(k)} ({\bf k})=-\lbrack A_{1,2}^{(0)} ({\bf
k})]^{-1}
\sum\limits_{m=0}^{s} A_{11}^{(m)} ({\bf k})m^{(k+m)} ({\bf k}) \\
M_{22}^{(k)} ({\bf k})=\lbrack A_{1,2}^{(0)} ({\bf k})]^{-2}
\sum\limits_{m=0}^{s} \sum\limits_{p=0}^{s} A_{11}^{(m)} ({\bf k})A_{11}^{(p)}
({\bf k})m^{(k+m+p)} ({\bf k})
\end{array}
\end{equation}       

We see that in order to calculate the matrices $A^{(m)} \;\lbrack 0\leq m\leq
s-1]$, we need to know the spectral moments $m^{(k)} ({\bf k})$ , relative to
the first field $\psi _{1} (x)$, for $0\leq k\leq 4s-1$. Lengthy calculations
show that the equation (2.6) can be expressed as
\begin{equation}
\sum\limits_{m=0}^{2s} a_{m} ({\bf k})m^{(k+m)} ({\bf k})=0\quad \quad \quad
\lbrack for\quad 0\leq k\leq 2s-1]
\end{equation}      
This result shows that the case $(n=2,s)$ is equivalent to the case
$(n=1,2s)$, when the special choice (4.1) is considered. We can make the
conjecture that for this particular choice the formulation depends only the
product $sn$. That is, two approximations characterized by the pairs $(n\/_{1}
,s\/_{1} )$ and $(n\/_{2} ,s\/_{2} )$ are equivalent if $n\/_{1} s\/_{1}
=n\/_{2} s\/_{2}$.

\section{The Hubbard model}

As an application of the formulation let us consider the Hubbard model,
described by the following Hamiltonian
\begin{equation}
H=\sum\limits_{ij} (t\/_{ij} -\mu \delta _{ij} )c^{\dagger}
(i)c(j)+U\sum\limits_{i} n_{\uparrow }(i)n_{\downarrow}(i)
\end{equation}      
The first term is the kinetic term which describes the motion of the electrons
among the sites of the Bravais lattice described by the vectors ${\bf R}_{i}
\quad [i=({\bf R}_{i} ,t)]$ ; $c(i),c^{\dagger} (i)$ are annihilation and
creation operators of electrons at site ${\bf R}_{i}$ in the spinor notation
\begin{equation}
c=\left( \begin{array}{c} c_{\uparrow} \\ c_{\downarrow} \end{array}\right) \\
\quad \quad \quad \quad c^{\dagger}=\left(c^{\dagger}_{\uparrow } \quad \quad
c^{\dagger}_{\downarrow}\right)
\end{equation}      
and satisfy canonical anti-commutation relations:
\begin{equation}
\{c_{\sigma}(i),c^{\dagger}_{\sigma'}(j)\}=\delta _{\sigma ,\sigma' } \delta
_{i,j} \quad \quad \quad \{c_{\sigma } (i),c_{\sigma' }
(j)\}=\{c_{\sigma}^{\dagger}
(i),c_{\sigma'}^{\dagger} (j)\}=0\end{equation}      
$\mu $ is the chemical potential. $t_{ij} $ denotes the transfer integral and
describes hopping between different sites. We fix the scale of the energy in
such a way that $t_{ii}=0$ . In momentum space the hopping matrix can be
written as
\begin{equation}
t_{ij} =\frac{1}{N} \sum\limits_{k}e^{i{\bf k}\cdot ({\bf R}\/_{i} -{\bf
R}\/_{j} )} t({\bf k})=\frac{\Omega }{(2\pi )^{d} } \int\limits_{\Omega _{B} }
d^{d} ke^{i{\bf k}\cdot ({\bf R}\/_{i} -{\bf R}\/_{j} )}
t({\bf k})\end{equation}      
where $N$ is the number of the sites, $d$ is the dimensionality of the system,
$\Omega $ is the volume of the unit cell in the coordinate space and
$\Omega_{B} $ is the volume of the Brillouin zone. In addition to the band
energy the model contains a term approximating the interaction among the
electrons. In the simplest form of the Hubbard model the interaction is
between electrons of opposite spin on the same lattice site; the strength of
the interaction is described by the parameter $U$ . $n_{\sigma } (i)=c_{\sigma
}^{\dagger} (i)c_{\sigma } (i)$ is the charge density of the electrons with
spin $\sigma $. The equation of motion for the electron field is
\begin{equation}
i\frac{\partial }{\partial t} c(i)=b(-i\vec{\nabla} )c(i)+U\eta
(i)\end{equation}      
where
\begin{equation}
\begin{tabular}{l}
$b(-i\vec{\nabla} )=-\mu +t(-i\vec{\nabla} ) $ \\
$\eta (i)=c^{\dagger} (i)c(i)c(i) $%
\end{tabular}
\end{equation}   

In the framework of the $sn$-pole approximation we shall consider the case
$n=2,\;s=2$ . Let us take as basis the two component field
\begin{equation}
\psi (i)=\left(
\begin{array}{c}
\psi _{1} (i) \\
\psi _{2} (i)
\end{array}
\right) \end{equation}      
and let us consider two possible choices.

\noindent {\bf I Choice}

\begin{equation}
\begin{tabular}{l}
$\psi _{1} (i)=c(i) $ \\
$\psi _{2} (i)=-\lbrack A_{1,2}^{(0)} (-i\vec{\nabla} )]^{-1}
\sum\limits_{m=0}^{2} A_{1,1}^{(m)} (-i\vec{\nabla} )\left( i\frac{\partial
}{%
\partial t} \right) ^{m} c(i) $%
\end{tabular}
\end{equation}       

This field satisfies the equation of motion
\begin{equation}
\left( i\frac{\partial }{\partial t} \right) ^{2} \psi
(i)=\sum\limits_{m=0}^{1} A^{(m)} (-i\vec{\nabla} )\left( i\frac{\partial
}{%
\partial t} \right) ^{m} \psi (i)+\delta j(i)
\end{equation} 
with the constrain
\begin{equation}
M^{(2+k)} ({\bf k})-\sum\limits_{m=0}^{1} A^{(m)} ({\bf k})M^{(m+k)} ({\bf
k})=0\quad \quad
\quad \lbrack 0\leq k\leq 1]\end{equation}      
As shown in Section 5, by taking this choice the formulation (n=2, s=2) is
equivalent to the case (n=1, s=4). Then, the retarded Green's function has the
expression
\begin{equation}
G({\bf k},\omega )=\sum\limits_{i=1}^{4} \frac{\sigma ^{(i)} ({\bf k})}{\omega
-E_{i}({\bf k})+i\eta }
\end{equation}      
The energy spectra are the roots of the equation
\begin{equation}
\sum\limits_{m=0}^{4} a_{m} ({\bf k})\omega ^{m} =0
\end{equation}      
with coefficients determined by the equations
\begin{equation}
\sum\limits_{m=0}^{4} a\/_{m} ({\bf k})m^{(k+m)} ({\bf k})=0\quad \quad \quad
\quad \lbrack 0\leq k\leq 3]
\end{equation}      

The spectral functions are
\begin{equation}
\sigma ^{(i)} ({\bf k})=\frac{1}{b^{(i)} ({\bf k})} \sum\limits_{m=0}^{3}
m^{(m)} ({\bf k})\sum\limits_{p=0}^{m} E_{i}^{p-m-1} ({\bf k})a_{p}
({\bf k})\end{equation}      
We need to calculate the first 8 electronic spectral moments
\begin{equation}
m^{({\bf k})} (k)=F.T.<\{\left( i\frac{\partial }{\partial t_{i} } \right)
^{k}c(i),c^{\dagger} (j)\}>_{E.T.} \end{equation}      
The calculation of the electronic spectral moments can be done by means of a
recurrence relation, as shown in Appendix D.

\noindent {\bf II Choice}

\begin{equation}
\psi (i)=\left(
\begin{array}{c}
\xi (i) \\
\eta (i)
\end{array}
\right) \quad \quad \quad \quad \quad
\begin{array}{l}
\xi (i)=\lbrack 1-c^{\dagger} (i)c(i)]c(i) \\
\eta (i)=c^{\dagger} (i)c(i)c(i)
\end{array}
\end{equation}      
This field satisfies the equation of motion
\begin{equation}
\left( i\frac{\partial }{\partial t} \right) ^{2} \psi
(x)-\sum\limits_{m=0}^{1} A^{(m)} (-i\vec{\nabla} )\left( i\frac{\partial
}{%
\partial t} \right) ^{m} \psi (x)=\delta j(x)
\end{equation}
with the constrain
\begin{equation}
M^{(2+k)} ({\bf k})-\sum\limits_{m=0}^{1} A^{(m)} ({\bf k})M^{(m+k)} ({\bf
k})=0\quad
\quad \quad \lbrack 0\leq k\leq 1]\end{equation}      
The spectral moments $M^{(k)} ({\bf k})=F.T.<\{\left(
i\partial\psi(i)/\partial t_{i} \right)^{k},\psi(j) \}_{t_{i}=t_{j}} >$ can be
expressed in terms of the electronic moments (6.15) by means of the following
relations
\begin{equation}
\begin{tabular}{l}
$M_{11}^{(k)} ({\bf k})=m^{(k)} ({\bf k})-2d^{(k)} ({\bf k})+g^{(k)}
({\bf k}) $ \\
$M_{12}^{(k)} ({\bf k})=M_{21}^{(k)} ({\bf k})=d^{(k)} ({\bf
k})-g^{(k)} ({\bf k}) $ \\
$M_{22}^{(k)} ({\bf k})=g^{(k)} ({\bf k}) $%
\end{tabular}
\end{equation}          
where
\begin{equation}
\begin{tabular}{l}
$d^{(k)} ({\bf k})=U^{-1} \lbrack m^{(k+1)} ({\bf k})-b({\bf
k})m^{(k)} ({\bf k})] $ \\
$g^{(k)} ({\bf k})=U^{-2} \lbrack m^{(k+2)} ({\bf k})-2b({\bf k})m^{(k+1)}
({\bf k})+b^{2}
({\bf k})m^{(k)} ({\bf k})] $%
\end{tabular}
\end{equation} 
We need to calculate the first 6 electronic spectral moments. The retarded
Green's function has a four-pole structure with energy spectra and spectral
functions calculated as shown in Section 5. As stated in Section II, we note
that the approximation has been made on the equation of motion for
second-order time derivative. No approximation has been done on the equations
for the first -order time derivatives. In particular, the following equations
\begin{equation}
\begin{tabular}{l}
$i\frac{\partial }{\partial t} \xi (i)=b(-i\vec{\nabla} )\xi (i)+t(-i\vec{\nabla} )\eta (i)-\pi (i) $ \\
$i\frac{\partial }{\partial t} \eta (i)=(U-\mu )\eta (i)+\pi (i) $%
\end{tabular}
\end{equation}       
where
\begin{equation}
\pi (i)=\frac{1}{2} \sigma ^{\mu } n_{\mu }(i)[t(-i\vec{\nabla}
)c(i)]+c(i)[t(-i\vec{\nabla})c^{\dagger}(i)]c(i)
\end{equation}      
are treated exactly in this approximation.

\section{Conclusions}
Among various analytical methods developed in the last decades to deal with
highly correlated electronic systems, the technique of expanding the Green's
function in terms of a finite number of spectral moments has been rather
successful in describing the properties of many correlated models. Along this
line of thinking, the crucial point is the choice of the basis in terms of
which the expansion is realized. A standard way is to choose the spectral
moments of the original fields. However, in presence of strong correlation
this choice may not provide the better basis for a perturbative approach and
other basis may be considered. In particular, a set of composite field
operators may be taken as the fundamental set in terms of which the Green's
function formulation is realized. In this process there are two ingredients
that should be considered: the choice of the basis and the dynamics. In this
work, we have presented a scheme of calculations where both aspects are taken
into account. Firstly, a set of $n$ composite basic fields is taken; the
choice of this set is made on the basis of physical criteria: the role played
by the interaction and by the value of the external parameters, the phase we
are interested to describe, and so on. Secondly, the dynamics is considered up
to the $s$-th order: we consider time derivatives of the fundamental fields of
increasing order and the truncation procedure is applied to the equation of
motion for the time derivative of order $s$, with the advantage that all
equations of motion up to order $s-1$ are treated exactly. Two theorems have
been proved about the conservation of the spectral moments in the approximation
considered. Finally, we have applied the formulation to the case of the
Hubbard models where some specific realizations have been considered. In
Appendix D we have presented a recurrence relation for the calculation of the
spectral moments relative to the Hubbard model.

\acknowledgements

The author is grateful to Dr. A. Avella for stimulating
discussions, for a very friendly collaboration and for his careful
reading of the manuscript. Valuable discussions with Prof. M.
Marinaro and R. M\"unzner are gratefully acknowledged.

\appendix

\section{Conservation of the first 2s spectral moments}
The spectral moments calculated in the $sn$ -pole approximation have the
expression
\begin{equation}
B^{(k)} ({\bf k})=\sum\limits_{i=1}^{sn} E_{i}^{k} ({\bf k})\sigma ^{(i)}
({\bf k})\end{equation}      
By using the expression (2.15) for the spectral functions, we have
\begin{equation}
B^{(k)} ({\bf k})=\sum\limits_{m=0}^{sn-1} \lambda ^{(m)} ({\bf k})P^{(k+m)}
({\bf k})\end{equation}      
where we have defined
\begin{equation}
P^{(k)} ({\bf k})=\sum\limits_{i=1}^{sn} \frac{E_{i}^{k} ({\bf k})}{b^{(i)}
({\bf k})}
\end{equation}      
By observing that the energy spectra satisfy the equation
\begin{equation}
\sum\limits_{m=0}^{sn} a_{m} ({\bf k})E_{i}^{m} ({\bf k})=0\quad \quad \quad
\quad
\lbrack for\quad i=1,\cdots sn]\end{equation}      
the quantities $P^{(k)} ({\bf k})$ satisfy the recurrence relation
\begin{equation}
\sum\limits_{m=0}^{sn} a_{m} ({\bf k})P^{(k+m)} ({\bf k})=0\quad \quad \quad
\quad
\lbrack for\quad k\geq 0]\end{equation}      
By simple algebraic relations we can show that
\begin{equation}
P^{(k)} ({\bf k})=\left\{
\begin{array}{l}
0\quad for\quad 0\leq k\leq sn-2 \\
1\quad for\quad k=sn-1
\end{array}
\right. \end{equation}      
Then, Eq.~(A2) can be written as
\begin{equation}
B^{(k)} ({\bf k})=\sum\limits_{m=sn-1-k}^{sn-1} \lambda ^{(m)} ({\bf
k})P^{(k+m)} ({\bf k})\quad \quad \quad \lbrack for\quad 0\leq k\leq
sn-1]\end{equation}      
When $0\leq k\leq 2s-1$ we have always $sn-1-k\geq sn-2s$ ; then, we can use
Eq.~(2.19) to obtain
\begin{equation}
\begin{tabular}{l}
$B^{(k)} ({\bf k})=\sum\limits_{m=sn-1-k}^{sn-1} \sum\limits_{p=0}^{sn-m-1}
M^{(sn-m-1-p)} ({\bf k})a_{sn-p} ({\bf k})P^{(k+m)} ({\bf k}) $ \\
\quad \quad \quad $=\sum\limits_{m=0}^{k} M^{(m)} ({\bf
k})\sum\limits_{p=sn-k+m}^{sn} a_{p} ({\bf k})P^{(p+k-1-m)}
({\bf k}) $ \\
\quad \quad \quad $=M^{(k)} ({\bf k})a_{sn} ({\bf k})P^{(sn-1)} ({\bf
k})+\sum\limits_{m=0}^{k-1} M^{(m)} ({\bf k})\sum\limits_{p=sn-k+m}^{sn} a_{p}
({\bf k})P^{(p+k-1-m)} ({\bf k}) $ \\
\quad \quad \quad $=M^{(k)} ({\bf k})\quad \quad \quad \quad \quad \quad \quad
\quad \quad \quad \quad \quad \quad \quad \quad \lbrack for\quad 0\leq k\leq
2s-1] $%
\end{tabular}
\end{equation}   
where Eq.~(A6) has been used. Then, we have the result that in the present
approximation the first $2s$ spectral moments are conserved. In closing this
appendix we note that from the above discussion the following relations follow

\noindent {\it (i) For $n=1$}

\begin{equation}
\begin{tabular}{l}
$B^{(k)} ({\bf k})=M^{(k)} ({\bf k})\quad \quad \quad \quad \quad \quad \quad
\quad
\quad for\quad 0\leq k\leq 2s-1 $ \\
$B^{(k)} ({\bf k})=-\sum\limits_{m=0}^{s-1} a_{m} ({\bf k})B^{(k-s+m)} ({\bf
k})\quad \quad
\quad \;for\quad k\geq s $%
\end{tabular}
\end{equation}   

\noindent {\it (ii) For $n\geq 2$}

\begin{equation}
\begin{array}{ll}
B^{(k)}({\bf k})=M^{(k)}({\bf k})\qquad &{\rm for}\ 0\leq k\leq 2s-1 \\
B^{(k)}({\bf k})=\sum\limits_{p=0}^{s-1} A^{(p)}({\bf k})B^{(k-s+p)}({\bf
k})\qquad &{\rm for}\ s\leq
k\leq sn-1  \\
B^{(k)}({\bf k})=-\sum\limits_{m=0}^{sn-1} a_{m}({\bf k})B^{(k-sn+m)}({\bf
k})\qquad &{\rm for}\ k\geq sn
\end{array}
\end{equation}       

\section{}
From the field equations (2.4) we have for $k\geq s$
\begin{equation}
\left( i\frac{\partial }{\partial t} \right) ^{k} \psi
(x)=\sum\limits_{m=0}^{s-1} A^{(m)} (-i\vec{\nabla} )\left( i\frac{\partial
}{\partial t} \right) ^{m+k-s} \psi (x)+\left( i\frac{\partial}{\partial t}
\right)^{k-s} \delta j^{(s)} (x)
\end{equation}      
Then, for $k\geq s$ the exact spectral moments satisfy the equation
\begin{equation}
M^{(k)} ({\bf k})=\sum\limits_{m=0}^{s-1} A^{(m)} ({\bf k})M^{(m+k-s)} ({\bf
k})+F^{(k-s)}
({\bf k}) \end{equation}      
where we defined
\begin{equation}
F^{(k)} ({\bf k})=F.T.<\left[ \left( i\frac{\partial }{\partial t} \right)^{k}
\delta j^{(s)} (x),\psi^{\dagger} (y) \right]_{\pm}>_{t_{x} =t_{y} }
\end{equation}      
Use of time translational invariance and of the constrains (2.5) leads to
\begin{equation}
F^{(k-s)} ({\bf k})=0\quad \quad \quad \lbrack for\quad s\leq k\leq
2s-1]\end{equation}      
Then, the spectral moments for $k\geq s$ can be written as
\begin{equation}
M^{(k)} ({\bf k})=\tilde{M} ^{(k)} ({\bf k})+\delta M^{(k)}
({\bf k})\end{equation}      
where $\tilde{M} ^{(k)} ({\bf k})$ and $\delta M^{(k)} ({\bf k})$ satisfy the
recurrence relations
\begin{equation}
\tilde{M} ^{(k)} ({\bf k})=\sum\limits_{m=0}^{s-1} A^{(m)} ({\bf k})\tilde{M}
^{(m+k-s)} ({\bf k})\quad \quad \quad \lbrack for\quad k\geq
s]\end{equation}      
\begin{equation}
\delta M^{(k)} ({\bf k})=F^{(k-s)} ({\bf k})+\sum\limits_{m=0}^{s-1} A^{(m)}
({\bf k})\delta M^{(m+k-s)} ({\bf k})\quad \quad \quad \lbrack for\quad k\geq
2s]\end{equation}      
with
\begin{equation}
\delta M^{(k)} ({\bf k})=0\quad \quad \quad \lbrack for\;0\leq k\leq
2s-1]\end{equation}      
We note that $\tilde{M} ^{(k)} ({\bf k})$ represents the spectral moment in
the approximation considered, when the higher-order field $\delta \/j^{(s)} $
is neglected. In Appendix C we prove the following: \\

\noindent {\bf Theorem V} Given $s$ matrices $A^{(m)} \;\lbrack 0\leq m\leq
s-1]$ of rank $n$, let us define the {\em characteristic} polynomial as
\begin{equation}
D(\omega )=Det\lbrack -\omega ^{s} +\sum\limits_{m=0}^{s-1} A^{(m)} \omega
^{m} ]=\sum\limits_{m=0}^{sn} \omega ^{m} a_{m}
\end{equation}      
and let us consider the $n \times n$ matrices $X^{(m)} \;\lbrack m\geq 0]$
determined by the following recurrence relation
\begin{equation}
X^{(k)} =\sum\limits_{m=0}^{s-1} A^{(m)} X^{(m+k-s)} \quad \quad \quad
\lbrack for\;k\geq s]\end{equation}      
Then, the following relation holds
\begin{equation}
\sum\limits_{m=0}^{sn} X^{(k+m)} a_{m} =0\quad \quad \quad \lbrack for\;k\geq
0]
\end{equation}      
for any choice of the matrices $X^{(m)} \;\lbrack 0\leq m\leq s-1]$. \\

By means of this theorem the matrices $\tilde{M} ^{(k)} ({\bf k})$ satisfy the
recurrence relation
\begin{equation}
\tilde{M} ^{(k)} ({\bf k})=-\sum\limits_{m=0}^{sn-1} a_{m} ({\bf k})\tilde{M}
^{(k-sn+m)} ({\bf k})\quad \quad \quad \lbrack for\;k\geq
sn]\end{equation}      

\section{Proof of Theorem V}

Given $s$ matrices $A^{(m)} \;\lbrack 0\leq m\leq s-1]$ of rank $n$, we
consider the matrix
\begin{equation}
P(\omega )=\omega ^{s} I-\sum\limits_{m=0}^{s-1} A^{(m)} \omega ^{m}
\end{equation}      
where $\omega $ is a parameter, which can generally be complex. $I$ is the
unit matrix in $n$ dimensions. Let us suppose that there exists the inverse
matrix
\begin{equation}
Q(\omega )=\lbrack \omega ^{s} I-\sum\limits_{m=0}^{s-1} A^{(m)} \omega ^{m}
]^{-1}
\end{equation}      
This matrix can be always written as a rational function with respect to the
parameter $\omega $
\begin{equation}
Q(\omega )=\frac{B(\omega )}{\varphi (\omega )}
\end{equation}      
The numerator $B(\omega )$ is a matrix whose elements are polynomials in
$\omega $ of degree $s(n-1)$ . The denominator $\varphi (\omega )$ is the
characteristic polynomial of degree $sn$
\begin{equation}
\varphi (\omega )=Det\lbrack -\omega ^{s} +\sum\limits_{m=0}^{s-1} A^{(m)}
\omega ^{m} ]=\sum\limits_{m=0}^{sn} \omega ^{m} a_{m}
\end{equation}      
Since $Q(\omega )$ is the inverse matrix of $P(\omega )$ we have
\begin{equation}
\lbrack \omega ^{s} I-\sum\limits_{m=0}^{s-1} A^{(m)} \omega ^{m} ]Q(\omega
)=I\end{equation}      
Use of Eq.~(C3) leads to
\begin{equation}
\lbrack \omega ^{s} I-\sum\limits_{m=0}^{s-1} A^{(m)} \omega ^{m} ]B(\omega
)=\varphi (\omega )I\end{equation}      

The two members of this identity are polynomials of degree $sn$ with matricial
coefficients. In order that this identity be satisfied, the coefficients for
each power $\omega ^{m} $ must be equal. In this identity we can replace the
variable $\omega ^{m} $ with a matrix $X^{(m+k)} \;\lbrack k\geq 0]$
\begin{equation}
\omega ^{m} \rightarrow X^{(m+k)}
\end{equation}      

If the matrices $X^{(m)} $ are chosen in such to satisfy the recurrence
relation
\begin{equation}
X^{(s+k)} =\sum\limits_{m=0}^{s-1} A^{(m)} X^{(m+k)} \quad \quad \quad
\lbrack for\;k\geq 0]\end{equation}      
then the left hand in (C6) vanishes, and so the second:
\begin{equation}
\varphi (X)=\sum\limits_{m=0}^{sn} a_{m} X^{(k+m)} =0\quad \quad \quad
\lbrack for\;k\geq 0]\end{equation}      
for any choice of the matrices $X^{(m)} \;\lbrack 0\leq m\leq s-1]$ .

\section{Spectral moments for the Hubbard model}

We recall the definition of the electronic spectral moments
\begin{equation}
m^{(k)} ({\bf k})=F.T.< \{ j^{(k)} (i),c^{} (j) \} >_{E.T.}
\end{equation}      
with
\begin{equation}
j^{(k)} (i)=\left( i\frac{\partial }{\partial t} \right) ^{k}
c(i)\end{equation}      
In order to calculate the anticommutators, we shall proceed in the following
way. From the Heisenberg equation (6.5) we obtain for $k\geq 1$
\begin{equation}
j^{(k)} (i)=b(-i\vec{\nabla} )j^{(k-1)} (i)+U\left( i\frac{\partial }{\partial
t}
\right) ^{k-1} \eta (i)\end{equation}      
with the convention that $j^{(0)} (i)=c(i)$ . By recalling the definition of
the Hubbard operator $\eta (i)$ [cfr. Eq.~(6.6)] we have
\begin{equation}
\left( i\frac{\partial }{\partial t} \right) ^{k} \eta
(i)=\sum\limits_{m=0}^{k} \left(
\begin{array}{c}
k \\
m
\end{array}
\right) \sum\limits_{l=0}^{m} \left(
\begin{array}{c}
m \\
l
\end{array}
\right) (-1)^{l} j_{\rho }^{(l) ^\dagger} (i)j_{\rho }^{(m-l)} (i)j^{(k-m)} (i)
\end{equation}      

The field equation (D3) can be written for $k\geq 1$ as
\begin{equation}
\begin{tabular}{l}
$j^{(k)} (i)=b(-i\vec{\nabla} )j^{(k-1)} (i) $ \\
\quad \quad \quad $+U\sum\limits_{m=0}^{k-1} \left(
\begin{array}{c}
k-1 \\
m
\end{array}
\right) \sum\limits_{l=0}^{m} \left(
\begin{array}{c}
m \\
l
\end{array}
\right) (-1)^{l} j_{\rho }^{(l) ^\dagger} (i)j_{\rho }^{(m-l)} (i)j^{(k-1-m)}
(i)$
\end{tabular}
\end{equation}      
We have a recurrence relation which allows us to calculate the source $j^{(k)}
(i)$. Let us define the quantities
\begin{equation}
\begin{tabular}{l}
$f_{\alpha \beta}^{k} (i,j)\;=\{ j_{\alpha }^{(k)} (i),c_{\beta
}^{\dagger} (j) \}_{E.T.} $ \\
$g_{\alpha \beta}^{k} (i,j)= \{ j_{\alpha }^{(k)^\dagger}
(i),c_{\beta}^{\dagger} (j) \}_{E.T.} $
\end{tabular}
\end{equation}      
By means of the previous results, these quantities satisfy the following
recurrence relations $for\quad k\geq 1$
\begin{eqnarray}
f_{\alpha \beta }^{k} (i,j)&=& b(-i\vec{\nabla} _{i} )f_{\alpha \beta }^{k-1}
(i,j) \nonumber \\
&& +U\sum\limits_{m=0}^{k-1} \left(
\begin{array}{c}
k-1 \\ m
\end{array}
\right) \sum\limits_{l=0}^{m} \left(
\begin{array}{c}
m \\ l
\end{array}
\right) (-1)^{l} \left[ j_{\rho}^{(l) ^\dagger} (i)j_{\rho }^{(m-l)}
(i)f_{\alpha \beta }^{k-1-m} (i,j) \right. \\
&&\left.-j_{\rho }^{(l) ^\dagger} (i)f_{\rho \beta }^{m-l} (i,j)j_{\alpha
}^{(k-1-m)} (i)+g_{\rho \beta }^{l} (i,j)j_{\rho }^{(m-l)} (i)j_{\alpha
}^{(k-1-m)} (i)\right]  \nonumber
\end{eqnarray}           
\begin{eqnarray}
g_{\alpha \beta }^{k} (i,j)&=& b(-i\vec{\nabla} _{i} )g_{\alpha \beta }^{k-1}
(i,j) \nonumber \\
&& +U\sum\limits_{m=0}^{k-1} \left(
\begin{array}{c}
k-1 \\ m
\end{array}
\right) \sum\limits_{l=0}^{m} \left(
\begin{array}{c}
m \\l
\end{array}
\right) (-1)^{l} \lbrack j_{\alpha }^{(k-1-m) ^\dagger} (i)j_{\rho }^{(m-l)
^\dagger} (i)f_{\rho \beta }^{l} (i,j) \\
&& -j_{\alpha }^{(k-1-m) ^\dagger} (i)g_{\rho \beta }^{m-l} (i,j)j_{\rho
}^{(l)} (i)+g_{\alpha \beta }^{k-1-m} (i,j)j_{\rho }^{(m-l) ^\dagger}
(i)j_{\rho }^{(l)} (i)] \nonumber
\end{eqnarray}    
with
\begin{equation}
\begin{tabular}{l}
$f_{\alpha \beta }^{0} (i,j)\;=\delta _{ij} \delta _{\alpha \beta } $
\\
$g_{\alpha \beta }^{0} (i,j)=0 $%
\end{tabular}
\end{equation}      

Then, the spectral moments are given by
\begin{equation}
m^{(k)} ({\bf k})=b({\bf k})m^{(k-1)} ({\bf k})+UL^{(k-1)} ({\bf k})\quad
\quad \quad \lbrack
for\quad k\geq 1]\end{equation}      
where
\begin{equation}
\begin{array}{l}
L_{\alpha \beta }^{(k-1)} ({\bf k})=\sum\limits_{m=0}^{k-1} \left(
\begin{array}{c}
k-1 \\
m
\end{array}
\right) \sum\limits_{l=0}^{m} \left(
\begin{array}{c}
m \\
l
\end{array}
\right) (-1)^{l} F.T.  \\
\quad \quad \quad \quad \left[ <j_{\rho }^{(l) ^\dagger} (i)j_{\rho }^{(m-l)}
(i)f_{\alpha \beta }^{k-1-m} (i,j)>_{E.T.} \right.\\
\quad \quad \quad \quad -<j_{\rho }^{(l) ^\dagger} (i)f_{\rho \beta }^{m-l}
(i,j)j_{\alpha }^{(k-1-m)} (i)>_{E.T.}  \\
\left.\quad \quad \quad \quad +<g_{\rho \beta }^{l} (i,j)j_{\rho }^{(m-l)}
(i)j_{\alpha }^{(k-1-m)} (i)>_{E.T.} \right]
\end{array}
\end{equation}      
with
\begin{equation}
m^{(0)} ({\bf k})=1
\end{equation}      

\end{document}